\documentclass[prc,showpacs,twocolumn,superscriptaddress]{revtex4-1}

\usepackage{graphicx}
\usepackage{longtable}
\usepackage{dcolumn}
\usepackage{epsfig}
\usepackage{lineno}

\begin{document}

\title{
Comparative study of the double $K$-shell-vacancy production \\
in single- and  double-electron capture decay
}%

\newcommand{\BNOINR}{Baksan Neutrino Observatory INR RAS, Neitrino 361609, Russia}
\newcommand{\KhNU}{V.N.~Karazin Kharkiv National University, Kharkiv 61022, Ukraine}
\newcommand{\Mendel}{D.I.~Mendeleev Institute for Metrology, Saint-Petersburg 190005, Russia}
\newcommand{\PNPI}{Petersburg Nuclear Physics Institute, NRC ``Kurchatov Institute'', Gatchina 188300, Russia}

\affiliation{\BNOINR}
\affiliation{\KhNU}
\affiliation{\Mendel}
\affiliation{\PNPI}

\author{S.S.~Ratkevich} \email{ssratkevich@karazin.ua} \affiliation{\BNOINR} \affiliation{\KhNU}
\author{A.M.~Gangapshev} \affiliation{\BNOINR}
\author{Yu.M.~Gavrilyuk} \affiliation{\BNOINR}
\author{F.F.~Karpeshin} \affiliation{\Mendel}
\author{V.V.~Kazalov} \affiliation{\BNOINR}
\author{V.V.~Kuzminov} \affiliation{\BNOINR}
\author{S.I.~Panasenko} \affiliation{\BNOINR} \affiliation{\KhNU}
\author{M.B.~Trzhaskovskaya} \affiliation{\PNPI}
\author{S.P.~Yakimenko} \affiliation{\BNOINR}

\date{\today}%

\begin{abstract}
\noindent
\textbf{Background:}
A double $K$-electron capture is a rare nuclear-atomic process in which two $K$ electrons are captured simultaneously from the atomic shell.
A ``hollow atom'' is created as a result of this process.
In single $K$ shell electron-capture decays, there is a small probability that the second electron in the $K$-shell is excited to an unoccupied level or can (mostly) be ejected to the continuum. In either case, a double vacancy is created in the $K$-shell.
The relaxation of the double $K$-shell vacancy,
accompanied by the emission of two $K$-fluorescence photons, makes it possible to perform experimental studies of such rare processes with the large-volume proportional gas chamber.
\\
\textbf{Purpose:}
The purpose of the present analysis is to estimate a double-$K$-shell vacancy creation probability per
$K$-shell electron capture $P_{KK}$ of $^{81}$Kr, as well as to measure the half-life of $^{78}$Kr relative to $2\nu2K$ capture.
\\
\textbf{Method:}
Time-resolving current pulse from the large low-background proportional counter (LPC), filled with the krypton sample, was applied to detect
triple coincidences of ``shaked'' electrons and two fluorescence photons.
\\
\textbf{Results:}
The number of $K$-shell vacancies per the $K$-electron capture, produced as a result of the shake-off process,
has been measured for the decay of $^{81}$Kr.
The probability for this decay was found to be $P_{KK}=(5.7\pm0.8)\times10^{-5}$ with a systematic error of
$(\Delta P_{KK})_{syst}=\pm0.4 \times10^{-5}$.
For the $^{78}{\rm Kr}(2\nu2K)$ decay, the comparative study of single- and double-capture decays allowed us to obtain the
signal-to-background ratio up to 15/1.
The half-life $T_{1/2}^{2\nu2K}(g.s. \rightarrow g.s.) = [1.9^{+1.3}_{-0.7}(stat)\pm0.3(syst)]\times 10^{22}$~y is determined from the
analysis of  data that have been accumulated over 782 days of live measurements in the experiment that used samples consisted of 170.6 g of
$^{78}$Kr.
\\
\textbf{Conclusions:}
The data collected during low background measurements using the LPC were analyzed to search the rare atomic and nuclear processes.
We have determined $P_{KK}^{exp}$ for the $EC$ decay of $^{81}$Kr, which are in satisfactory agreement with $Z^{-2}$ dependence of $P_{KK}$
predicted by Primakoff and Porter.
This made possible to more accurately determine the background contribution in the energy region of our interest for the search for the
$2K$-capture in $^{78}$Kr.
The general procedure of data analysis allowed us to determine the half-life of $^{78}$Kr relative to $2\nu2K$-transition with a greater
statistical accuracy than in our previous works.

\end{abstract}


\maketitle


\section{\label{Intr}Introduction}

A double electron capture ($ECEC$) is a process inverse to a double beta decay.
This process can occur in atoms on the proton-rich side of the mass parabola of even-even isobars.
In this reaction two bound electrons from the atomic shell are captured by two protons, thereby lowering the charge of the final nucleus by two units:
\begin{equation}
  e^{-}_b + e^{-}_b + (Z,A) \rightarrow  (Z-2,A)^{**} (+2\nu_e).  \label{ecec}
\end{equation}
Here, the two asterisks denote the possibility of leaving the system in an excited nuclear and/or atomic state, the latter being characterized by two vacancies in the electron shell of the neutral atom.
The two-neutrino mode $(2\nu ECEC)$ is allowed in the Standard
Model while the existence of the lepton number violating
neutrinoless double electron capture $(0\nu ECEC)$ would prove the Majorana nature of the neutrino.
Unfortunately, the search for $ECEC$ decay is complicated due to small rates and the experimental challenge to observe only the produced X-rays or Auger electrons for transitions to the ground state.
It is less probable than  a double beta decay with the
emission of two electrons  and
two neutrinos ($2\nu\beta \beta$).
The latter process has been detected in direct experiments for 11 different nuclei \cite{bb_nndc_bnl,Bar2015NP}, whereas there are only a few indications of the $2\nu ECEC$ process which have been seen in $^{78}$Kr. They were discussed in our previous works
\cite{Gavr2013PRC,PAN2013}, and a geochemical measurement for  $^{130}$Ba \cite{Meshik:2001,Pujol:2009} with the half-lives of the order of
$10^{21}$ years.
However, from the experimental registration prospective it is beneficial to keep looking for the $2\nu 2K$ process.
The objective of these searches would be a triple coincidence detection of two $K$ X-rays of fluorescence generated by the electron occupation of two formed vacancies, and of low-energy Auger electrons from the residual excitation.
The distinct feature of these X-rays is that they differ from the regular fluorescence by energy and intensity.
Moreover, there is a considerable difference in the nature of the fluorescence when comparing neutrinoless vs two neutrino double electron
capture processes.

In the present work, a two-neutrino double $K$-electron capture ($2\nu 2K$) is another process of our interest.
The analysis of the investigated nuclei showed that the capture of two electrons from the $K$-shell provides the largest contribution to
the process of $ECEC$.
The data reported in Ref.~\cite{Doi1992} suggest that the $2K$-capture events for $^{78}$Kr and $^{124}$Xe
isotopes account for 78.6\%
and 76.8\%
of the total number of $ECEC$ processes.

Apparently, a unique state is formed in a daughter atom when two electrons from the $K$ shell are captured.
This state represents a neutral atom with the inflated shell, exposing two vacancies in the $K$ shell.
In order to detect such a process, we have to keep in mind that for $Z > 30$, where $K$-fluorescence yields are large, the dominant decay of
double $K$ vacancy states happens through the
sequential emission of two $K$ X-rays.

Our search for the $2K$-capture in  $^{78}$Kr and $^{124}$Xe isotopes have been carried out in a deep underground low background
laboratory of the Baksan Neutrino Observatory (BNO) for several years \cite{Gavr98,Gavr2011,Gavr15Xe124,Gavr17Xe124}.
Large copper proportional counters (LPC)
with the full volume of 10.4 l were employed there at the current stage of research.
This choice of counters was dictated by their reliability and simplicity of design, especially when operating in the average X-ray energy field
that is smaller than 100 keV.
A detailed description of the setup can be found, for example, in Ref.~\cite{PTE}.
The technique of searches for the $2K$-capture process was based on a comparison of pulse-height spectra in the range of total energy
relaxation of an excited electron shell of the atom.
For this aim, we used the data of measurements from samples with a different isotopic enrichment.

The purpose of the present work is
to establish a time correlation (coincidence) between the two $K$ X-rays that are emitted after the decay of double $K$-vacancy states.
Comparative analysis of the signals, coming from the creation of double $K$-shell-vacancy states, was carried out.
These signals are produced by the $2\nu2K$-capture of the $^{78}$Kr or when one $K$ electron is captured and the other one is ejected from the
atom in the $^{81}$Kr.

In the first approximation, one may consider that an additional vacancy does not affect the
fluorescence yield.
In our  first works, we assumed that the fluorescence yields for
($K^{-2}\rightarrow K^{-1}L^{-1}$) and ($K^{-1}L^{-1}\rightarrow L^{-2}$) transition are equal to the fluorescence yield of normal diagram line
($K^{-1}\rightarrow L^{-1}$).
As a result, we were able to approximately calculate the following:
the total energy ($B_{2K}$ - energy of the double $K$-electron), the
composition and the energy of radiation
(a cascade of two characteristic photons and Auger electrons),
the
probability to fill the double $K$ vacancy with the emission of two photons of given characteristic energies (the second degree coefficient of
the fluorescence yield $\omega_{2K}\simeq\omega_{K}^2$).

A characteristic X-ray can travel a long distance in a gas medium from the point of its origin to the point of its absorption.
For example, 10\%
of $K$ X-rays with energies of 11.2 and 12.5 keV are absorbed in krypton at the pressure of 4.4 Bar ($\rho=0.0164$~g/cm$^3$) on paths
of 1.8 and 2.4 mm long, respectively.
The paths of electrons with the same energies are 0.37 and 0.44 mm, respectively.
They produce almost pointwise charge clusters of a primary ionization in the gas \cite{Drift and diffusion}.
In the case when both X-rays and  Auger electrons are absorbed in the gas of the proportional counter operating
volume, the total
energy deposition is distributed
over three pointwise charge clusters $(X \otimes X \otimes  e_{A})$.
These events possess a number of unique features.

For this goal, we can make a comparative analysis of the selected spectra of three-point events. Arranging the point energy deposits in these events in order of magnitude, we can select only those that contain a small component of the several keV in range of energy deposits, the ratio between two others being higher than 0.7. The average and maximum amplitudes are generated by characteristic $K$ photons of tens of keV
$([K_{\alpha1} \otimes K_{\alpha1}],
[K_{\alpha1} \otimes K_{\alpha2}],
[K_{\alpha2} \otimes K_{\alpha2}],
[K_{\alpha1} \otimes K_{\beta1}],
[K_{\alpha1} \otimes K_{\beta2}],
[K_{\alpha2} \otimes K_{\beta1}]$,
and $[K_{\alpha2} \otimes K_{\beta2}]$
of the daughter atom. The fraction $\alpha _k$ of the listed combinations of two quanta detected within a three-point event is equal to 0.985 out of the total number of possible combinations. Selecting three-point events on the basis of this criterion thus allows us to additionally reduce the background within the energy range of the expected peak and to obtain a general idea of the spectrum's shape over a wide range of energies.

Such an approach enabled us to establish a certain limit on the half-life of the considered decay, even when no essential effects was
observed.
The situation had changed when in our long-continued measurements appeared a positive effect ($\sim 10$ events/year) in the range of released energy $B_{2K}$ in data with an exposure of more than a hundred kg$\times$d  \cite{PAN2013,Gavr17Xe124}.
This had happened as a result of numerous successive improvements of the methods of data analysis and background characteristics of the experimental setup.

To interpret this excess correctly, precise values of the mentioned above parameters
[$B_{2K}$, $\omega_{2K}$, $E(K_{\alpha,\beta})$, $I(K_{\alpha,\beta})$] are essential.
These values can only be calculated in an appropriate
theoretical model.
This model has to describe processes that occur in the excited atomic shell.
Below, we suppose, that the fluorescence yields are proportional to the energy in third power.
In this approximation, we recalculate the above-listed parameters.
The related radiative widths and energies are calculate within the framework of the Dirac-Fock method.
The corresponding high-statistics and high-resolution experiment is necessary to test such calculations for the $2K$-capture.
It will be challenging to perform such an experiment, though.

It is known that, besides the $2K$-capture, other rare physical phenomena that can create a double $K$ vacancy in atoms exist
\cite{Freedman_74}.
For example, a double $K$-shell photoionization of the atom can create the ``hollow atom'' by absorbing a single photon and releasing both $K$
electrons.
Another example is studies of double $K$ ionization that follows electron capture $(EC)$ decay of radioactive nuclei.
During this process, there is a small probability that the second electron in the $K$-shell is excited to the unoccupied level
(shake-up; SU) or ejected to the continuum (shake-off; SO).
These processes will appear as a background in $2K$-capture experiments, because gamma-rays are absorbed there within the fiducial volume of
the counter. These gamma-rays are coming from the walls of the detector.
In addition, the long-lived cosmogenic radionuclide $^{81}$Kr in the depleted krypton sample is present in residual amounts.
$^{81}$Kr decays by $K$ electron capture can create two vacancies in the $K$-shell of a daughter atom.

Atoms with an empty $K$ shell can not be metastable.
An electron relaxation and rearrangement processes can give us a response to a $K$-shell doubly excited state.
The excited atom decays in a cascade of nonradiative Auger and radiative transitions.
The most probable $K$ hypersatellite transitions
are those for which one of the two $K$ vacancies is filled by a $L$-electron, namely $KK-KLX$ ($X = L, M$, etc.) hypersatellites in Auger electron spectra and $KK-KL$ (usually noted $K_\alpha ^h$) hypersatellites in X-ray fluorescence spectra.
For the experimental method used by us, radiative transitions
are more preferable.

\begin{figure}
\includegraphics*[width=1.05in,angle=270.]{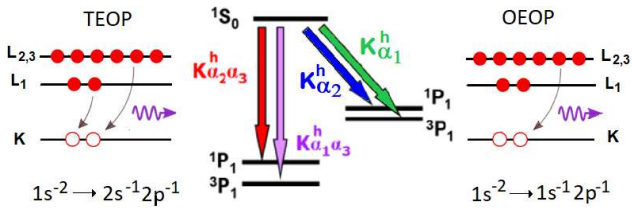}
\caption{\label{diagr_fig}
Schematic of the TEOP (left) and OEOP (right) transitions and the atomic level decay diagram for the
initial $K$-shell two-hole state $^1S_0$ (middle).
}
\end{figure}

The radiative decay of double $1s$ vacancy states may proceed through two electron
one-photon (TEOP) or one-electron one-photon (OEOP) transitions \cite{Heisenberg,Baptista84}.
The scheme and the atomic level decay diagram for these transitions with the fluorescence of the $ K_\alpha $-series are shown in Fig.~\ref{diagr_fig}.
In the TEOP $(1s^{-2}\rightarrow 2s^{-1}2p^{-1})$  transition, which is denoted by  $K_{\alpha \alpha}^h$,
the two $K$-shell core holes are filled simultaneously via a correlated two-electron jump.
Moreover, there is a single photon emitted.
Its energy is approximately twice as large as that of the emitted parent $K_\alpha$ diagram line.
The OEOP process, which corresponds to the $K_\alpha^h$ $(1s^{-2}\rightarrow 1s^{-1}2p^{-1})$  hypersatellite transition,
represents to a large extent the predominant radiative decay channel of hollow $K$-shell atoms.

Two electron one photon transitions have been observed in heavy-ion-atom collision experiments since the seventies.
The distinct feature of TEOP transitions is that they are characterized by extremely weak intensities as compared to those of
hypersatellites.
It is because they correspond to correlated multielectron processes.
To date, there are several theoretical and experimental works (see Ref.~\cite{Mukherjee2009} and references therein) whose purpose is to study
two electron one photon processes from doubly ionized atoms; the ionization is created by the inner shells.

Recently, the first experimental evidence of TEOP transitions in the single-photon $K$-shell double ionization was reported by Hoszowska
\emph{et al.} \cite{HoszowskaPRL2011}.
The $K_{\alpha \alpha}^h$ to $K_{\alpha}^h$  branching ratios for Mg, Al, and Si were found to be of the order of $10^{-4}$.
To our best knowledge, there has been only one experimental observation of $K_{\alpha\alpha}^h$ X-rays in the electron-capture decay.
The Mn $K_{\alpha\alpha}^h$ X-rays were observed by Yasuhito Isozumi in a measurement done with a Si(Li) detector \cite{Isozumi_PRA80}.
The intensity ratio of $K_{\alpha \alpha}^h$ X-rays to $K_{\alpha}$ X-ray was found to be $R=(5.5 \pm 1.2) \times 10^{-8}$ for the $EC$ decay
of $^{55}$Fe.
Thus, the contribution of TEOP transitions in our measurements can be considered negligible.

The radiative decay of double $K$-shell hole states proceeds mainly through the OEOP process.
In this case, the second radiative transition $(1s^{-1}2p^{-1}\rightarrow 2p^{-2})$ arising when, in addition to a single $K$ vacancy, there are one or more additional vacancies in the outer orbitals of the initial state will be called a $K^s$ satellite.
Thus, double $K$-shell vacancies give rise to the subsequent emission of hypersatellite $K$ X-rays and satellite $K$ X-rays.
In our case, a coincident detection of these X-rays is the main criterium for the selection of events due to the OEOP process, and it allows us
to determine the production probability of the double vacancy in $K$-shell.

In this paper, we focus on the analysis of theoretical and experimental data accumulated by the processes of formation of hollow atoms
as a result of the single $K$-electron capture in $^{81}$Kr and double $K$-electron capture decay in $^{78}$Kr.
The paper is structured as follows. In Sec.~\ref{sec2} we discuss features of the fluorescence generated by a two-neutrino double $K$-capture
for $^{78}$Kr and $^{124}$Xe.
Section~\ref{sec3} describes the experimental procedures and analysis of data obtained from the measurements done with the LPC which were
filled with two samples of the krypton.
The conclusion is given in Sec.~\ref{sec4}.

\section{\label{sec2}Calculation}

The $ECEC$ process is possible when the mass difference $Q$ between the parent and daughter atoms is
positive:
\begin{equation}
\label{eq:Q}
Q=M(Z,A)-M(Z-2,A),
\end{equation}
where $M(Z,A)$ is a mass of the atom with the atomic number $Z$ and the mass number $A$.

In principle, the total energy of the transition can be carried away by two emitted
neutrinos.
This mechanism is possible only for the higher orders of perturbation theory since it requires a radiationless rearrangement of the electron
shell.
In order to obtain a clear and complete picture, we consider two mechanisms.
One mechanism includes the capture of two valence electrons. As a result of this, the daughter atom is left in the ground state.
The probability of such a mechanism existing is relatively low, due to the low probability of finding both outer electrons on the nucleus.
The other possible mechanism involves the
capture of two inner electrons, followed by the simultaneous radiationless transition of
two valence electrons to the formed vacancies.
This is due to the non-orthogonality of the electron wave functions of $(Z, A)$ and $(Z-2, A)$ atoms  \cite{Karp4,Karp5}.
The probability of this to happen is also low due to the low amplitude of the overlap between the wave functions of inner and outer electrons.
For example, the overlapping integral of the wave function of the
$4s$ electron of the krypton atom and the $1s$ electron of the selenium atom is $3\times10^{-3}$. Hence, the amplitude of the
simultaneous transition of two electrons is around $\sim 10^{-5}$.
This means that the probability of such mechanism to happen is as low as $\sim 10^{-10}$.

\subsubsection{Widths of fluorescent bands upon $2\nu 2K$-capture}

Let us consider a conventional mechanism of the double electron capture that is observed in the lowest order of perturbation theory.
This mechanism involves the nuclear capture of two $K$ electrons.
As a result of this capture, there is an atom, with an excited shell, formed.
In addition, if the energy conservation allows, there is a nuclear excitation formed, as well.
We shall assume that the excitation energy of the electron shell after the emission of the
two neutrinos is $\varepsilon_A$, and that of the nucleus, is $\varepsilon_n$. We denote the mass of
the atom with two vacancies available as $M^{**}_{Z-2}\equiv M_{Z-2}+\varepsilon_A$, and
that of the atom with one $K$-vacancy which becomes available after the emission of the first fluorescence quantum, as $M^{*}_{Z-2}$.
Moreover,
each atom remains in the excited state after the emission of two fluorescence quanta.
We denote the masses of (mentioned above) excited states as $M^{f}_{Z-2}$.
Consequently, by using Eq.~(\ref{eq:Q}) the energy-release for the $ECEC$ process can be written as
\begin{equation}
\label{eq:Q2}
Q^*=M_Z-M_{Z-2}-\varepsilon_A-\varepsilon_n \equiv M_Z-M^{**}_{Z-2} -\varepsilon_n.
\end{equation}
Let us consider the transition of a nucleus to the ground state, which is characterized by $\varepsilon_n=0$.
This transition is accompanied by the energy release
\begin{equation}
\label{eq:Q3}
Q_1=M_Z-M^*_{Z-1}.
\end{equation}
Ordinarily,$ Q_1\ll 0$. The first intermediate state of the nucleus
$(Z - 1, A)$ is virtual, with the descent from the mass shell
(defect of the resonance)
\begin{equation}
\label{eq:Q4}
\Delta _1=M_Z-\varepsilon^1_\nu-M^*_{Z-1} = Q^*_1-\varepsilon^1_\nu\ll 0.
\end{equation}
Here and below, $\varepsilon^i_\nu$ and $\varepsilon^i_X$ represent the energy of the
$i^{\rm{th}}$ neutrino and of the fluorescence quantum,
respectively.
Similarly, the defect of the resonance for the second intermediate state,
\begin{equation}
\label{eq:Q5}
\Delta_2=Q^*-\varepsilon^1_\nu-\varepsilon^2_\nu.
\end{equation}
For the third propagator,
\begin{equation}
\label{eq=Q6}
\Delta_3=M_Z-M^*_{Z-2}-\varepsilon^1_\nu-\varepsilon^2_\nu-\varepsilon^1_X,
\end{equation}
and for the fourth,
\begin{equation}
\label{eq=Q7}
\Delta_4=Q_f-\varepsilon^1_\nu-\varepsilon^2_\nu-\varepsilon^1_X-\varepsilon^2_X,
\end{equation}
$Q_f=M_Z-M^f_{Z-2}$ is the final energy release that happens between the initial and the final state.

The formed vacancies are filled by the emission of two fluorescence quanta so
that the portion of the released energy $Q^*$ (\ref{eq:Q2}) is
carried away by four particles: two neutrinos and two characteristic $K$ photons.

As follows from Ref.~\cite{Karp1},
the energy distribution of the fluorescence quanta is
\begin{widetext}
\begin{eqnarray}\label{eq:Q13}
W_{2\nu 2e} \sim \Gamma^1_\nu \Gamma^2_\nu
\frac{\Gamma^1_X \Gamma^2_X}{[(\varepsilon^1_X-\Omega_1)^2+(\frac{\Gamma_2+\Gamma_3}{2})^2]
[(\varepsilon^2_X-\Omega_2)^2+(\frac{\Gamma_f+\Gamma_2}{2})^2]}
\frac{d\varepsilon^1_X}{2\pi}\frac{d\varepsilon^2_X}{2\pi}.
\end{eqnarray}
\end{widetext}
Here, $\Gamma^i_\nu$ and $\Gamma^i_X$ are the widths corresponding to the elementary
amplitudes of the electron capture and fluorescence, respectively.

Expression (\ref{eq:Q13}) describes the shape of the fluorescence spectrum.
As expected, the shape of this spectrum has two peaks with resonance energies given by the difference between the full
energy of the atom before and after the fluorescence:
\begin{equation}
\label{eq:Q14}
\Omega_1=M^*_{Z-2}-M_{Z-2}-\varepsilon_A,
\end{equation}
\begin{equation}
\label{eq:Q15}
\Omega_2=\varepsilon_A-\varepsilon_f.
\end{equation}
As it was mentioned above, these resonance energies differ from the regular values that are observed in the case of single-hole
decay states.
Moreover, the peaks are considerably wider.
$\Gamma_2$ in Eq. (\ref{eq:Q13}) is the full width of the decay of the atomic state with two vacancies in the $K$-shell per time unit.
Considering that the probability of the fluorescence is proportional to the number of
vacancies, we may assume that its estimate would be approximately double the total width of
the atom with
one vacancy on the $K$ shell, $\Gamma_K$:
\begin{equation}
\label{eq:Q16}
\Gamma_2\approx 2\Gamma_K.
\end{equation}
After the initial fluorescence, the hole is usually transferred to the $L$-shell, so that the atom remains in the state with two vacancies.
Its total width $\Gamma_3$, defined in Eq. (\ref{eq:Q13}) is approximately
\begin{equation}
\label{eq:Q17}
\Gamma_3 \approx  \Gamma_K +\Gamma_L \approx 2\Gamma_K,
\end{equation}
where $\Gamma_L$ is the width of the hole state in the $L$ shell. In
the first approximation, this width is equal to the width
of a vacancy in the $K$ shell \cite{Karp9}. Finally, $\Gamma_f$ is the width
of the atom in the state following two fluorescence
with two vacancies in the $L$ shell.
Similarly to previous estimations, it can be approximated to
\begin{equation}
\label{eq:Q18}
\Gamma_f \approx  2\Gamma_K.
\end{equation}
Combining all the estimates, we find from Eq. (\ref{eq:Q13}) that the
width of the first fluorescence quantum is approximately $\Gamma_2
+ \Gamma_3 \approx 4 \Gamma_K$, as well as the width of the second
fluorescence quantum is approximately $\Gamma_3 + \Gamma_f \approx
4 \Gamma_K$. Therefore, one can assume that widths of the
fluorescence lines for $2\nu ECEC$ are quite wide - on the order of
four widths of the regular fluorescence.

\subsubsection{Fluorescence energy upon $2\nu 2K$-capture}

A unique state of a neutral atom with an inflated electron shell and unfilled $K$-shell is formed as a result of the double $K$-capture.
A number of questions arise:

--- What energy $E^{**}_{Z-2}$ is needed to inflate the atomic shell, and what energy release is expected upon the $2K$-capture?

In the first approximation, the energy $E^{**}_{Z-2} = 2K_{ab}$, where $K_{ab}$ is the energy required to transfer the $K$-electron to the
valence level.
More accurate calculation of $E^{**}_{Z-2}$ can be performed if one remembers that the value of $E^{**}$ is given by the energy difference
between ``inflated'' and ground states of the atom.
Such a calculation was performed numerically in Ref.~\cite{Karp10} by using the RAINE computer code package  \cite{RAINE_89,RAINE_93}.
This package calculates atomic shell parameters by using multiconfigurational Dirac-Fock (MCDF) method.
The finite size of the nucleus, the highest
quantum-electrodynamic correction for vacuum polarization, and the
electron self energy were considered there. As a result, a value of
$E^{**}_{\rm Se} = 25804$ eV for selenium and a value of $E^{**}_{\rm Te}
= 64390$ eV for tellurium were obtained. The remaining energy,
$E_3=E^{**}_{Z-2}-\varepsilon^1_X-\varepsilon^2_X$, is released by the atom
in the form of Auger electrons and soft X-rays, which are emitted after the emission of
two fluorescence photons.

---  What are the expected energies of the characteristic $K$ photons?

In the first approximation,
energies of the characteristic $K$ photons are equal to those of ordinary fluorescence
quanta $K^0_{\alpha,\beta}$. More accurate
values were calculated using the RAINE package.
We calculated the energies of $K^h_{\alpha1,2}$ hypersatellite and $K^s_{\alpha1,2}$ satellite transitions for selenium and tellurium.
The first fluorescence quantum differs most from the energy of $K^0_{\alpha1}$, with the energy difference being equal to $\sim360$ and
$\sim680$
eV for selenium and tellurium, respectively.
The second quantum is larger than the ordinary fluorescence by about 60 and 165 eV for selenium and tellurium, respectively.
The results of our calculation are presented in Table \ref{tab:Calculated-and-reference}.
\begin{table*}
\renewcommand{\arraystretch}{1.3}\fontsize{10}{12}\selectfont\setlength{\tabcolsep}{0.4em}
\caption{
Calculated  values of photon energies (in eV), released energy ($B_{2K}$, in eV) and fluorescence yields in comparison with table values.
\label{tab:Calculated-and-reference}
}
\begin{ruledtabular}
\begin{tabular} {ccccccccccccc}
 atom & \multicolumn{3}{c}{Referenced data}      & chain & \multicolumn{8}{c}{Present calculation}   \\
 \cline{1-1}                  \cline{2-4}        \cline{5-5}                              \cline{6-13}
  ~  & $K_{\alpha1}^{0}$ & $\omega$ & $2B_K$ & ~    & $K_{\alpha1}$ & $K_{\alpha1}^{h}$ & $ K_{\alpha2}^{h}$ & $K_{\alpha1}^{s}$ & $
  K_{\alpha2}^{s}$ & $B_{2K}$ & $\omega^h$ & $\omega^s$ \\
\hline
Se   & 11222 & 0.602 & 25304 & \ref{rec1a} & 11248 & 11606 & 11560 &   11308 & 11262 &25804 & 0.770 & 0.610 \\
Br   & 11924 & 0.636 & 26948 & \ref{rec_K} & 11932 & ~     & ~     &   ~     & ~     &~     & ~     & ~     \\
 ~   & ~     & ~     & ~     & \ref{rec_KK}& ~     & 12294 & ~     &   11984 & ~     &27528 & 0.648 & 0.636 \\
Te   & 27472 & 0.875 & 63620 & \ref{rec1a} & 27471 & 28151 & 27858 &   27636 & 27362 &64390 & 0.883 & 0.874 \\
\end{tabular}
\end{ruledtabular}
\end{table*}
The corresponding values were calculated
based on the energies of the electron separation given in Ref.~\cite{x-ray}.

--- What is the fluorescence yield of two $K$ photons?

Vacancies in the $K$-shell are filled mostly by the fluorescence.
The emission of Auger electrons is also possible. The probability
of a radiative transfer is doubled upon the availability of two
vacancies, in compliance with the number of vacancies in the initial
state.
The probability of the Auger transfer to happen is also doubled.
The probability of radiative transfer over a unit of time diminishes by 1/6 for the second fluorescence quantum.
This is proportional to the occupation number of the $P$-shell in its initial state.
The probability of the second Auger transfer to happen falls off over time  by exactly the same
value; hence, the expected probabilities of the fluorescence upon the
$K$-capture remain the same as we would expect in the first approximation,
for example, 60.2\%
for selenium
and 87.5\%
for tellurium
\cite{x-ray}.

One can estimate corrections to the fluorescence yields by taking into account the changes of the electron shell.
In particular, we can employ reference values of the normal X-rays.
This is assured by the fact that the change of the electronic configuration leads to the change of X-rays energies.
Corresponding data are presented in Table~\ref{tab:Calculated-and-reference}.
Radiative widths there increase with the energy as $\sim k^3$ \cite{Ah}.
The Auger process can be considered as a conversion process \cite{bk}, whose energy dependence is much weaker and, thus, it may be neglected in
the first approximation.
We also neglect the changes induced by changing electronic wave functions  \cite{land}.
Let us now consider a simple two-channel model, which assumes that the probability of the emission of an X-ray reads as the branching ratio
\begin{equation}
\omega_X = \frac{\Gamma_X}{\Gamma_X+\Gamma_A}\,, \label{br}
\end{equation}
where $\Gamma_X$ and $\Gamma_A$ are radiative and Auger partial  widths, respectively.
In accordance with our assumption, let the modified radiative widths be
\begin{equation}
\Gamma_X' = (k'/k)^3\Gamma_X \equiv (1+\xi)\Gamma_X\,,  \label{k3}
\end{equation}
where $k$ and $k'$ are the table and actual values of the transition energy.
By substituting Eq.(\ref{k3}) in Eq.(\ref{br}) and taking into account the fact that $\xi\ll 1$, one gets the following expression
\begin{equation}
\omega_X'=\frac{\Gamma_X'}{\Gamma_X'+\Gamma_A}=\omega_X[1+\xi(1-\omega_X)]\,.
\label{br1}
\end{equation}
We present in Table~\ref{tab:Calculated-and-reference} the modified values of the
fluorescence yields, calculated by means of Eq.(\ref{br1}).
As one can see, these changes are within one percent of each other, as it was expected.
They are in an agreement with the numerical calculation \cite{2K-25}, where changes of the electron shell were explicitly taken into account.

\section{\label{sec3}EXPERIMENTAL PROCEDURES}

\subsection{Introduction}

For the LPC, the signal from the $2\nu 2K$-capture is similar to the signal from the double-vacancy creation that follows the single electron
capture.
In both cases, two simultaneous fluorescence photons are emitted from the daughter atom,
\\
\\
\emph{chain associated with the double $K$-capture}
\begin{eqnarray}
  2e_K + (Z,A) &\rightarrow & (Z-2,A)^{K^{++}} +2\nu_e,  \label{rec1a}
  \\  & & \hookrightarrow {(Z-2,A)} + K^h_{\alpha} + K^s_{\alpha} + m\times e_ A,
  \nonumber
\end{eqnarray}
\emph{chain associated with single $K$-capture}
\begin{eqnarray}
  e_K+(Z,A) &\rightarrow & (Z-1,A)^{K^{+}} + \nu_e \nonumber\\
    & & \hookrightarrow (Z-1,A) + K_{\alpha} + e_A   \label{rec_K} \\
    e_K+(Z,A) &\rightarrow & (Z-1,A)^{K^{++}}  + \nu_e + e_K  \nonumber\\
    & & \hookrightarrow (Z-1,A) + K^h_{\alpha} + K^s_{\alpha} + e_A.  \label{rec_KK}
\end{eqnarray}
The total energy release and radiation characteristics of SO processes
in the single $K$-capture decay of $^{81}$Kr are close to
the ones of the $2K$-capture in $^{78}$Kr.
This means that they can create a background for the decay under investigation.

The results of our calculations, within the overall theoretical
model, can be tested by the analysis of the data collected in the calibration mode of the detector
with the sample containing $^{81}$Kr.

The total probability of $K$-shell SO and SU processes to occur is on the order of
$10^{-4}$ per a single $K$-capture.
Compensatory effects of the sudden reduction of a nuclear charge and sudden disappearance of electron-electron Coulomb interaction originate
the mentioned SO and SU processes.
The contribution from the SU process to this probability is negligible \cite{Campbell1991}.
The filling of the empty $K$-shell is accompanied by the nearly simultaneous
($\sim10^{-16}$ s) emission of either two $K$ X-rays and two Auger electrons, or one $K$ X-ray
and one Auger electron.

In order to find the probability of the double-$K$-vacancy production that follows the $K$-shell electron capture, data accumulated during
long-term experiments were used.
The purpose of these measurements was to search for the $2K$-capture of $^{78}$Kr and
resonance absorption of the solar axion emitted in the $M1$ transition of $^{83}$Kr nuclei \cite{JETPLet2015}.

The measurement of the background was performed using two samples of krypton with different isotope composition: one of  enriched in $^{78}$Kr
up to 99.81\%
and the second one is a depleted krypton, left after the $^{78}$Kr isotope extraction up to 0.002\%.
These samples had been deliberately cleared from  the radioactive isotope of $^{85}$Kr ($T_{1/2} = 10.756$ y), which is present in the
atmospheric krypton.
The procedure of purification of enriched sample was described
in Ref. \cite{Kr-78purf}.
Table~I in Ref.~\cite{Gavr2013PRC} gives final isotopic compositions of the krypton samples.

The cosmogenic radioactive isotope of $^{81}$Kr with the volume activity of  $(0.076 \pm 0.004)$ min$^{-1}$l$^{-1}$~Kr \cite{Kr81atm} was
contained
in the original atmospheric krypton.
The significant part of this isotope comes into the sample of depleted krypton during the process of its production.
$^{81}$Kr produces $^{81}$Br by decaying via the electron capture.

The residual activity is sufficient enough to study the process of shake-off in the daughter $^{81}$Br atom.
The source activity was on the order of 4 $EC$ decays/min.
The sample of krypton, enriched in $^{78}$Kr, has been cleaned up from isotope $^{81}$Kr.

\subsection{Experimental runs}

Three stages of measurements were performed with the krypton sample containing $^{81}$Kr with the total exposure of 1167 live days distributed
evenly throughout the search period.
The total collection time of the first and second stages was 7600 and 8400 h, respectively.
The LPC operated at the pressure of 4.4 and 4.6 bar during in the first and second stages, respectively.
The total time of the third stage amounted to 12,000 h, with a pressure of 5.6 bar.
The total number of recorded $EC$ decays had reached $6.7\times 10^6$ events.

Two stages of long-lived measurements were carried out with krypton enriched in $^{78}$Kr.
In this case, the LPC operated at a pressure of 4.4 and 4.6 bar at the first and second stage, respectively, as with the depleted sample.
This data set has also been analyzed in Refs.\cite{Gavr2013PRC,PAN2013}.

The ultimate sensitivity of the setup to the target effects depends first on the value of the intrinsic detector background and the quality of the method used to isolate the desired signal, with the specified feature set, from this background.
Therefore, measurements were carried out in an underground laboratory
of the BNO at 4900 m w.e. depth \cite{GAVR_NIMA13}.
The flux of muons there was $(2.6 \pm 0.09)$ m$^{-2}$ day$^{-1}$ \cite{Gavrin_prnt}.
The LPC was surrounded by a passive shield made of copper (20 cm), lead (20 cm), and polyethylene (8 cm).

The signal from the anode of the counter was delivered to the charge-sensitive amplifier (CSA) connected
to the high-voltage electrode through the capacitor.
Raw pulse shapes from the CSA were digitized with a high sampling rate and analyzed offline by using directly the pulse shape discrimination
technique.
CSA parameters had been optimized for the transmission of a signal
with minimum distortions.
In addition, the information about the primary ionization charge spatial distribution, projected onto the radius of the counter, was fully
described by the shape of the pulse.

Collected data were processed offline by applying the optimum filtering procedure \cite{PTE}.
Besides this, the following pulse-characterizing parameters were evaluated for
every signal that was recorded:
the pulse amplitude, the rise time,
and several pulse-shape indicators (the afterpulse amplitude, the delay time afterpulse).
In addition, the energy resolution of the filtered baseline noise and the amplitude of
the signal for a given energy deposit were estimated for each of data sets (100 hours of measurements).

Output charge pulses from coincidence of two or more photon events from the CSA look like the pileup signals.
There is a mathematical procedure that allows to convert the registered charge pulse into the pulse of the electric current, produced by
electrons of primary ionization up to the boundary of the avalanche region.
The obtained shape of the current's pulse can be described by a set of Gaussian curves that enable us to determine the charge brought in a
certain component of the multipoint event.
Calculated area of an individual Gaussian should correspond to the charge
(energy) of the corresponding pointlike ionization.
The digitized pulses' processing methods used in the experiment were described in the Ref.\cite{Ratk06}.

Hypersatellite and satellite transitions of a certain energy are the signatures of the presence of double $K$-vacancy states.
As it was noted in the introduction, characteristic X-rays and  Auger electrons can produce almost pointwise charge clusters of primary
ionization
in the gas under pressure that was used in the experiment.
Depending on the pulse rise times of registered events, these events can be classified as ``point-like'' or ``lengthy''.
Point-like pulses are produced by particles that create primary
ionization charges in a region with a characteristic length (along with the radius) that is smaller than the cathode radius.
Long-run, complex events can be produced when several point-like events are registered simultaneously at different distances from the anode
(so-called ``multi-point'' or ``multi-cluster'' events).
Besides that, extended pulses can be generated by high-speed electrons with a track length comparable to the counter's radius.

The pulse shape of registered charged particles and photons that is taken from the LPC has some specific features.
This fact enables us to easily distinguish the useful pulses from the non-ionized noise signals and
microdischarge events. The pulse taken from the anode wire is formed mainly due to the negative charge (ionic component) induced onto the
anode
by positive ions.
These ions are produced near the anode wire during the gas amplification process, and they move towards the cathode.
The drift time of ionic component is $\sim 0.447$ s.
The contribution from the charge that is induced by avalanches of electrons and that is acting on the anode is small enough ($\sim 7$\%).
This happens because electrons have to pass a small distance on their way to the anode (electrons collection time is $\sim 1$ ns).

The electric current pulse shape is affected by  the density distribution of primary ionization electrons crossing the border of the gas
amplification region.
Parameters of the distribution  depend on the drift time of the originally pointlike charge that drifts towards the anode.

The pointlike ionization is smeared out during the drift time due to the diffusion of the electrons into a cloud with a radial charge density
distribution.
This distribution appears to be similar to the normal distribution.
Besides that, the CSA discharge constant $\tau _{o}$ affects the pulse shape, as well.
Therefore, as mentioned above, the output pulses CSA were converted to current pulses.

\begin{figure}
\includegraphics*[width=2.5in,angle=0.]{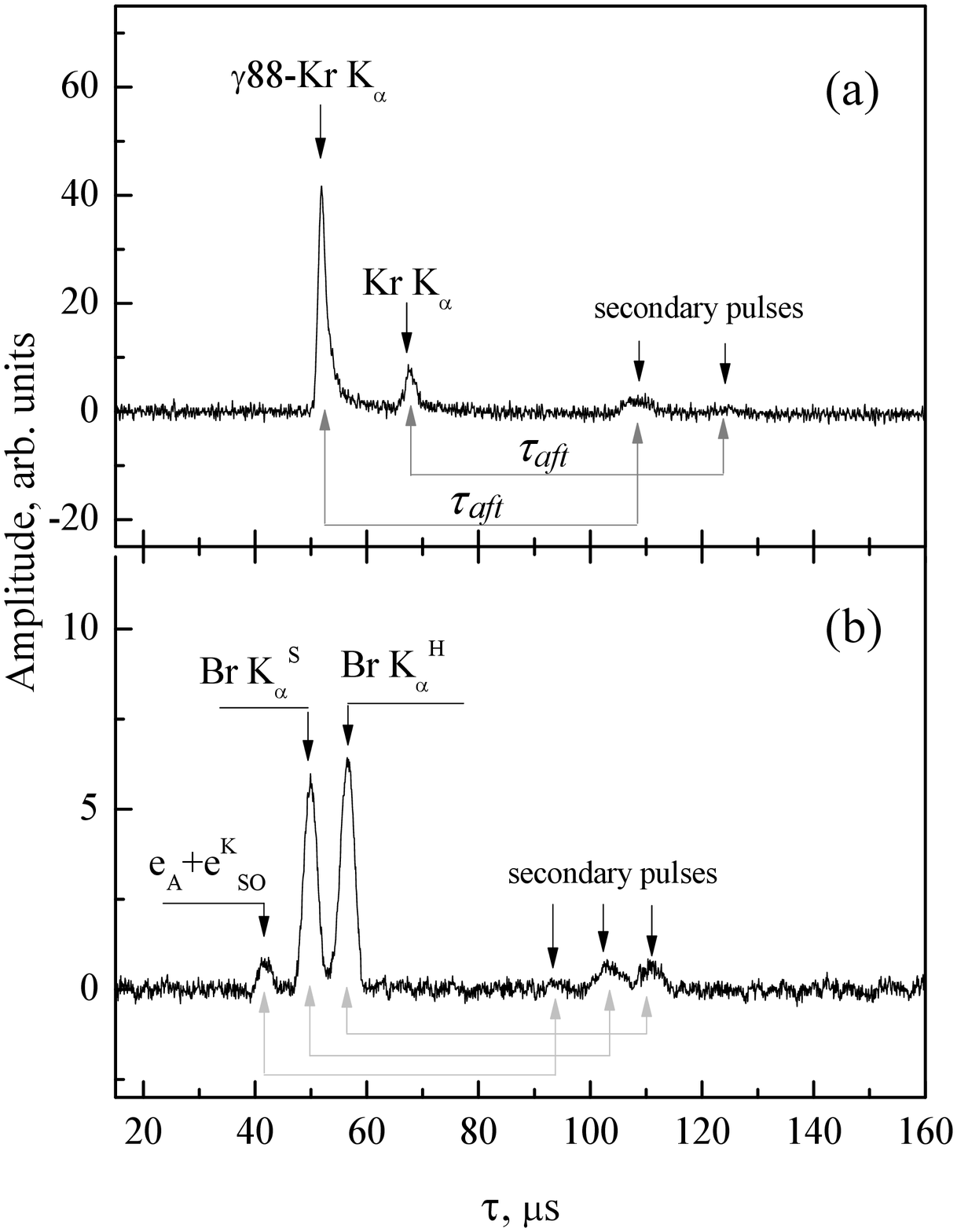}
\caption{\label{impulse_point_23}
Examples of curent pulses of two types of events: (a) - a two-point event for photoabsorption of an 88 keV photon with simultaneous
emission of a 12.6 keV characteristic photon Kr $K_\alpha$
and a photoelectron $E_{{\rm p.e.}}=\gamma 88 - K_\alpha({\rm Kr})=75.4$ keV; (b) - three-point event with simultaneous emission of two
characteristic
X-ray and electrons.
$\tau _{aft}$ is the duration of the electron drift from the cathode to the anode.
}
\end{figure}

The time resolution of subpulses of the current LPC signal allows us to implement a technique that studies coincidence events of Auger
electrons and X-rays in the offline mode.
It will be shown below that, despite a poor energy resolution
(compared to monolithic germanium or silicon drift detectors), the measurement of weak signals is possible. These signals are characterized by
energy signatures that differ slightly from a random background.
Examples of current pulses corresponding to events of two types are presented in Fig.~\ref{impulse_point_23}.
The pulse (the differential of the charge pulse) due to the photoabsorption of a 88~keV photon with simultaneous emission of characteristic
photon
with an energy of 12.6 keV and a photoelectron $E_{{\rm p.e.}} = \gamma 88 - K_\alpha({\rm Kr}) = 75.4$ keV
(a two-point event of $[K_\alpha \oplus {\rm p.e.}]$) is shown in Fig.\ref{impulse_point_23}(a).
As it can be seen from Fig.~\ref{impulse_point_23}(a) the current pulse
is not of a symmetric shape.
It contains ionic and electronic components with different drift times in the LPC gas.
The output current pulse can be transformed to a symmetric shape.
For this purpose, the correction on a self-discharge of the CSA has been made and
the electronic component of the output pulse is used only.

\subsection{Data analysis}

The procedure that we used to determine the electron component
of pointlike primary-ionization clusters via the
analysis of the entire shape of the LPC pulse was discussed in detail in Ref.~\cite{PTE}.
Ibid,  wavelet thresholding methods of noise removal from the current pulses have been described too.

The electric current signal candidate for the three-point event from the $(X \otimes X \otimes e_{A})$ coincidence is illustrated in
Fig.\ref{impulse_point_23} (b).
The maximum distance between pointwise charge clusters in projection onto the counter's radius is equal to the radius.
The cathode-to-anode drift time of electrons (that is the maximum collecting time of electrons) is of the order of $\sim53$~$\mu$s for the pure
krypton \cite{PTE}.
From Fig.\ref{impulse_point_23}, it is apparent that the secondary pulse was produced in the counter $\sim (53-54)$~$\mu$s
after the first pulse. The amplitude of the secondary pulse (afterpulse) is about five to six times smaller.
These secondary pulses have nothing to the incident radiation
but are generated from secondary processes that arise from effects within the primary
avalanches. These spurious pulses could lead to the multiple counting, while only one pulse should be recorded.
The optical radiation emitted by the excitation of the electron shell of an atom in a Townsend avalanche causes the release of the
photoelectrons
from the cathode \cite{Campion73}.
Since the working gas contains no quenchers, the probability of the photoeffect on the cathode is rather high.
Besides the first afterpulse, there are many others.
Their amplitudes are damped increasingly.
As it can be seen, the initial
pulse and the afterpulse are of a similar shape but are different in duration.
The afterpulse of a longer duration can be explained by the diffusion
expansion of the photoelectron cloud during its drift from the cathode to the
anode.

The relative magnitude of the afterpulse depends on the coordinate of the event along the anode wire.
Furthermore, the gas amplification factor is reduced significantly near its edges.
There was no gas amplification in these segments, and charges were collected in the ionization mode.
As was shown in Ref.~\cite{Gavr2010}, such events may be excluded from the data by introducing a discrimination in the relative coordinate
($\lambda$) of events along the anode wire. The coordinate
$\lambda$ is determined as the ratio of the first-afterpulse
amplitude to the pulse amplitude.
The relative afterpulse amplitude depends on the solid angle at which the
cathode surface is seen from the anode wire. The angle in question is maximal at the anode's midpoint in length, and it is minimal at the
anode's edge. The value of $\lambda$ changes accordingly.
For real pulses of our counter, this parameter $\lambda$ has to be in the following range: $0.05 \leq \lambda \leq 0.40$.
Discarding pulses for which $\lambda < \lambda_k$, where $\lambda_k$ is maximum value for the end events, one can fully eliminate edge events.
In this case, there will be a small loss of useful events in the three-point spectrum (the coefficient of selection of useful events is
$k_\lambda = 0.89\pm0.02$).

\begin{figure}
\includegraphics*[width=2.25in,angle=0.]{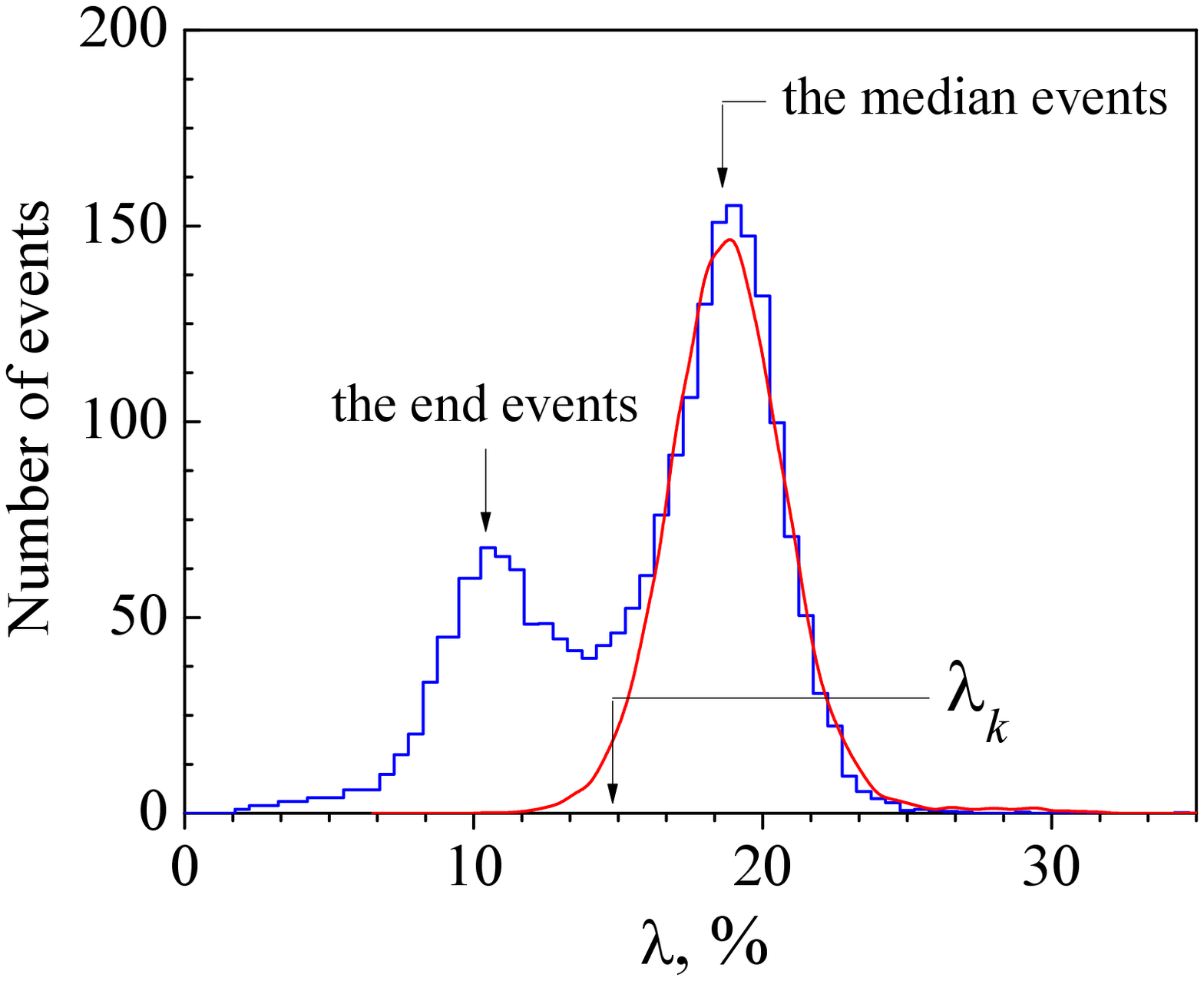}
\caption{\label{spcB_Krn_II}
The distributions of the events in the LPC filled with depleted
krypton with energies 10-16 keV ($^{81}$Kr, red curve)
and the background events of detector  in the energy region 20-80 keV (blue histogram).}
\end{figure}

As an example, Fig.~\ref{spcB_Krn_II}
shows the lambda distribution of two-point events with the energy range of 20-80 keV for the background measurements of
the LPC that is filled with the depleted krypton during the second stage of measurements (blue histogram).
The red curve is the $\lambda$ distributions of events of energies 10-16 keV, and the main contribution there comes from the $K$-capture of
$^{81}$Kr.
Thus, we can select the event in the counter with the same gas gain and thereby improve energy resolution.

Despite the fact that counters are sealed, their lifetime is limited by a microscopic degasation  of inner material.
This leads to the gradual contamination of the filling gas.
The influence of electronegative impurities is considerably pronounced in large-volume counters and counters with small values of electron
drift velocity.
The quenching and accelerating gases enter the working gas during the long-term operation of the counter. Presumably, this happens due to the
slow sublimation of the polyatomic gas molecule coming from the surfaces of casing and insulators.
For example, oxygen gradually diffuses from the sealing ring of the flanges
LPC and it acts as a quenching additive.
This leads to a gradual change in the detector's performance.
Such parameters of the detector as the duration of the electron drift from the cathode to the anode $\tau _{aft}$, the parameter $\lambda _k$,
the energy resolution and the gas gain varied with the time of event registration.
These facts had to be taken into account for the purposes of our experiment.

The timestamp of event occurrence allows us to monitor changes in the energy calibration and the aforementioned operating parameters of the
detector during long measurements.
This made it possible to improve the energy resolution via an adapted digital
signal processing.
For this purpose, regular calibration of the detector by the external $^{109}$Cd-source was done biweekly \cite {PTE}.
The sample that contained signatures of the radioactive isotope $^{81}$Kr enabled us to study the operating characteristics of the detector in
more detail.

\begin{figure*}
\includegraphics*[width=3.55in,angle=270.]{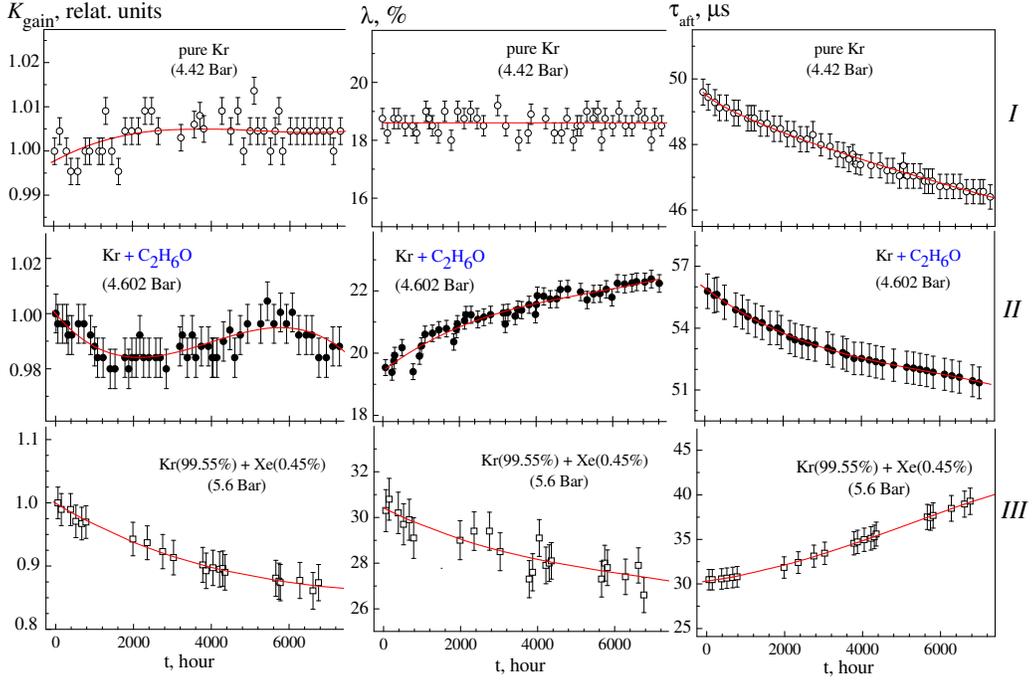}%
\caption{\label{calibr_Kr_nat_I_II_III}
Variations of relative coefficient gas gain ($K_{gain}$)) and
the centroid $\lambda^c(t)$- and $ \tau^c_{aft}(t)$-distribution versus time
in each of the three stages (\emph{I}, \emph{II} and \emph{III}) of measurement sample depleted krypton.}
\end{figure*}

\begin{figure*}
\includegraphics*[width=1.65in,angle=270.]{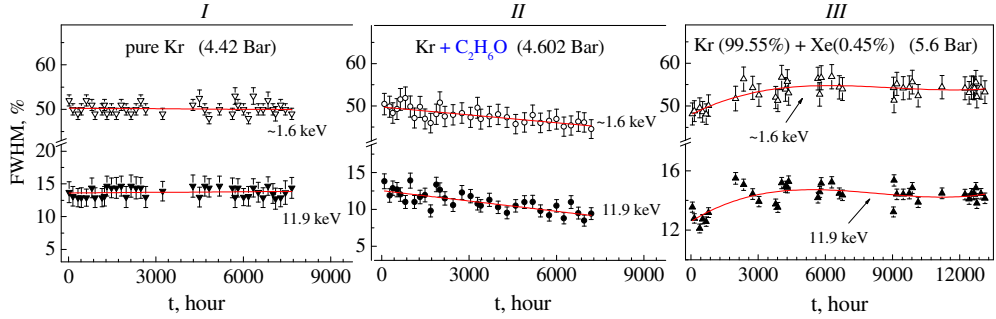}%
\caption{\label{FWHM_I_II_III_run}
Variations of energy resolution FWHM of LPC versus time in (\emph{I}, \emph{II} and \emph{III}) stage, respectively.}
\end{figure*}

Results of a determination of variations of spectrometric parameters of the LPC for samples of pure krypton, krypton with trace contamination
of the polyatomic molecules, and krypton with the  0.45\%
of Xe as a quenching gas are presented in Figs.~\ref{calibr_Kr_nat_I_II_III} and \ref{FWHM_I_II_III_run}.
The relative change in the gas gain ($K_{gain}$) and centroid $\lambda^c (t)$ - and $\tau^c _{aft} (t)$-distributions (accumulated over 100 h),
recorded over three long stages of accumulation of events in the counter, filled with a depleted sample, are shown in
Fig.~\ref{calibr_Kr_nat_I_II_III}.
Time dependence of the energy resolution [full width at the half maximum (FWHM)] of the LPC for same three stages is shown in
Fig.~\ref{FWHM_I_II_III_run}.
As it can be seen, there, operating parameters of the detector for all three stages of measurements behaved differently with time.
Therefore, a permanent correction of the response function of the detector was required.
As an example, we needed to deal with ethyl alcohol and acetone which were used to clean the contaminants during assembling the detector.
As a result, organic molecules got into the working gas during the long-term operation of the counter due to the slow sublimation from the
surfaces of the casing and insulators of the counter.
Polyatomic gas molecules can help to retain the counter proportionality by suppressing the avalanche breeding, which would otherwise occur when
the excited inert gas atoms emit  ultraviolet photons.

Events that have been accumulated over the course of measurements may be split into groups, depending on the number of ionization clusters
(pointwise charge clusters) in a single event. In an ideal case of a complete separation, all events that are initiated by charged particles,
end up in the group of one-point events (of a single continuous trajectory). Events that are associated with a photon absorption are
distributed among the groups of one- and two-point events in accordance with the probabilities of the Compton process, photoabsorption, and the
probability for the characteristic radiation yield upon the photoabsorption induced filling of the $K$-vacancy in the target atom.

Two-point events take place if, for example, 1) a photon undergoes the Compton scattering and it is absorbed in the counter's volume,
2) a photon undergoes photoeffect and X-ray are emitted in the process of the vacancy filling,
3) an electron creates the bremsstrahlung.
In all these cases, secondary photon is absorbed at some distance from
its origin and creates second cluster of ionization. If these two
clusters are located at a different distance from the anode, the detected
signal will be made of two pulses,
with the time difference corresponding to
the moments of entering the gas amplification region by two groups
of primary electrons.

Three-point events can appear if all secondary radiation is absorbed
inside the counter, as in, 1) Compton scattering of a photon, followed by the photoeffect and
release of the
X-ray by filling the vacancy in the atomic shell,
2) photoeffect followed by the $K$-shell ionization, by the photoelectron in another
atom and release of two X-rays.
Three-point events, which are the main goal of our study, have to be  formed as a result of absorption  into the fiducial
volume of the counter of two characteristic photons
and the Auger electrons generated by the double $K$-shell vacancy production in the investigated krypton samples.

All more-than-three-point events are considered as multipoint events.
They can appear, for example, as a result of Compton scattering of
the photon by the $K$-electron, followed by the $K$-shell photoeffect and release of two $K$ X-rays.
Events can move around within multipoit groups because of possible overlaps of the pulses of separate components of multi-point events.
In addition, it can also be caused by the noise effect.

The response function of the LPC was determined using the radioactive point source of
$^{109}$Cd, which is characterized by the 88 keV line with 0.036 yields per
decay.
Photons passed through the collimating hole in the copper shield, which is located below the middle point of the counter.
Then, corresponding spectra were compared with the results of Monte Carlo simulations.
The effective dimensions of the detection
setup were adjusted to obtain a satisfactory agreement between
the measured and simulated spectra.
This enabled us to calculate the response of the detector to radiation of arbitrary energy
within the range of our interest.

The Monte Carlo simulation of the experiment was carried out
in order to determine the detection yields for $K$-shell shake-off
electrons after $K$-capture and the influence of background
events such as an internal bremsstrahlung, $K$-, $L$-, and $M$-shell
electron ejection.

Taking into account the composition of events from the $^{81}$Kr and $^{109}$Cd sources and results of the simulation of these processes in
the
LPC,
one can determine the quality of the suppression of the noise component
and the efficiency of the procedure for separating events based on the multipoint event criteria.
We simulated the electron-photon transport inside the detector by using the Monte-Carlo method implemented in the user code, embedded into
a universal package GEANT4~\cite{GEANT4} with the model of Low Energy Electromagnetic Package called Penelope 2008 \cite{PENELOPE}.

The Low Energy Electromagnetic package implements a precise treatment of electromagnetic
interactions of photons and electrons with matter for energies of a few hundred eV. This package includes a component that takes into account
the process of atomic relaxation.
It simulates the processes of de-excitation of the atom, which is left in the excited state  by the creation of a vacancy originated by primary
particles (photons, electrons),
and the emission of X-ray fluorescence and Auger electrons.
The calculation of the efficiency
of the selection of multipoint events uses reference values of fluorescence
yields and absorption coefficients.

\subsection{Double $K$-shell ionization following K-capture in $^{81}$Kr}

\begin{figure}
\begin{centering}
\includegraphics*[width=2.75in,angle=0.]{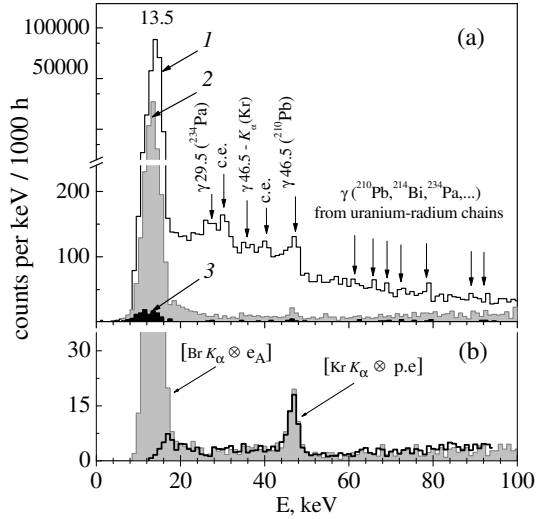}
\par\end{centering}
\caption{\label{fig:SpcE_81Kr}
Top panel - the pulse height spectra  of  (1) single-,  (2) two-, and (3) three-point events that form the total analyzed pulse amplitude
spectrum of the counter detecting both background and $^{81}$Kr decays.
Bottom panel -the pulse height spectrum of the two-point events with one subpulses in the energy range
$E=12.6 \pm 1.9$ keV for the two filling samples of LPC: with krypton enriched in $^{78}$Kr (dark line) and with
depleted krypton containing $^{81}$Kr (shaded region).
}
\end{figure}

Figure \ref{fig:SpcE_81Kr} shows the pulse-height spectra of (1) single-~, (2) two-~, and (3) three-point events of the background of the LPC
filled with depleted sample gas.
The spectra were obtained using the procedure for correcting the response function taking into account the time dependence.
The spectra are normalized to 1000 h of data that were accumulated during the second stage of the experiment.
Low-intensity lines from radioactive contaminants that resided on the walls of the counter can be seen clearly on the top of the histogram.

\begin{figure}
\begin{centering}
\includegraphics*[width=1.5in,angle=0.]{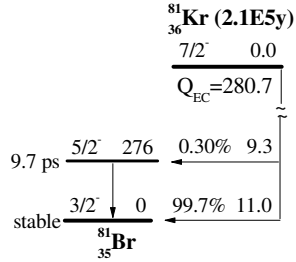}
\par\end{centering}
\caption{\label{fig:81Kr_diagr}
The decay scheme of $^{81}$Kr.
}
\end{figure}

In Fig.~\ref{fig:SpcE_81Kr}, one can see a clear peak that corresponds to the energy of 13.5 keV.
The energy resolution of the peak is ${\rm FWHM} \approx 15.3$\%.
This peak occurs when the $K$-electron capture of $^{81}$Kr happens.
The decay scheme of $^{81}$Kr is shown in Fig.~\ref{fig:81Kr_diagr}.

The $7/2^+$ ground-state nucleus of $^{81}$Kr decays mainly (99.70\%)
into the $3/2^-$ ground state of $^{81}$Br via the electron capture
with a half-life of $T_{1/2} = 2.3\times 10^5$ yr \cite{Baglin2008}.
This transition is of the ``first-forbidden unique'' type.
It can only occur during the electron capture process
because the energy $Q_{EC}$ of this decay is $(280.8 \pm 0.5)$ keV \cite{Baglin2008}.
Alternatively, the mentioned nucleus can decay into the excited $5/2^-$ state of $^{81}$Br with a subsequent emission of the 276 keV photon,
which  occurs with the probability of $3.0(2) \times  10^{-3}$ \cite{Axelsson88}.
This transition is of a first-forbidden non unique type.
The decay energy of this rare process equals only 4.7 keV and it is much smaller than the
$K$-electron binding energy in a bromine atom, which amounts
to $B_K = 13.47$ keV. Therefore, any transition to the excited state that happens in $^{81}$Br  proceeds only by electron captures from higher
shells. Consequently, the $K$-electron capture can occur only during a
ground-state to ground-state transition.
$K$-shell electron capture occurs with a probability of $P_K = 0.875$ \cite{Chew74}.

The lines that are shown in Fig.~\ref{fig:SpcE_81Kr} and that correspond to the energy range above 18 keV are identified as emitted by members
of uranium-radium and
actinium natural radioactivity chains, starting from $^{238}$U and $^{235}$U, respectively.
For example, the intense $\gamma$-line at 46.5 keV is emitted by $^{210}$Bi after the $\beta$-decay of $^{210}$Pb ($T_{1/2}=22.2$ yrs).
The $^{210}$Pb isotope
is produced after $^{222}$Rn ($T_{1/2}=3.82$ days, $\alpha$-decay) in the $^{238}$U decay chain. It could either be produced directly in the
copper, due to decays of internal radioactive micro impurities (volume source), or, it could have been accumulated on the casing surface
(surface source) during the detector's manufacturing stage, due to the deposition of daughter products coming from the decay of atmospheric
$^{222}$Rn.
$\gamma$-radiation comes into the fiducial volume of the counter from both sources. There are mainly single-point events (due to the
photoeffect on the atom of krypton with the deexcitation of Auger-electrons) and two-point events (due to the photoeffect with an emission of
the krypton characteristic radiation) that correspond to the peak line of 46.5 keV.
Only a small portion of events can be considered as three-point ones. These are the events for which a primary photon or a characteristic
photon scatters off outer-shell electrons with a subsequent absorption of a secondary quantum.

As a result of the beta-decay of $^{234}$Th,  which is being a decay product of $^{238}$U,  63.3, 73.9, 92.4 and 92.8 keV lines are emitted by
$^{234}$Pa. The 63.3 keV line coincides with the $E2$ transition line of 29.5 keV, which has a large $L$-shell conversion coefficient in the
protactinium.
Another $L$-converted transition of the 20.0 keV line $(M1 + E2)$ in $^{234}$Pa can also be seen in the spectrum of two-point events.

There are peaks in the spectrum at $\sim 28$, $\sim 31$ keV and $\sim 43$ keV that could be associated with conversion electrons (c.e.).
C.e. can be originated from the surface source, $^{210}$Pb.
The most intense lines are those of 30.1 keV (52\% per decay)
and 43.3 keV (13.6\% per decay).
These peaks should be represented by single-point events.
The observed shift (with respect to the expected line) of a maximum energy, with respect to the expected line, could be explained by the
deposit from  a portion of near-wall primary ionization electrons on the cathode due to their diffusion.
With the emission of conversion electrons, the residual excitation
of the $^{210}$Bi daughter shell relaxes by radiation of
characteristic photons of $L$-series ($E_{L_\alpha}=10.8$ keV,
9.3\%
per decay; $E_{L_\beta}=13.0$ keV, 11.2\%
per decay and etc.)
or/and by Auger electrons. In different combinations, this
radiation can enter the fiducial volume simultaneously with
$\beta$-particles and c.e. In the simplest case, when a c.e.
of 30.2 kev energy and a characteristic $L_\alpha$-photon are registered,
a two point event with the energy deposit of $\sim42.5$ keV is
produced in the gas.  The fraction of such events is small enough
in the total spectrum.

The photoelectric effect is the dominant for the energy range of our interest.
Interaction cross
sections for other competing mechanisms, such as the Compton scattering or
incoherent scattering, are about thousand times smaller.
The photoelectric effect
involves the interaction of a photon with a tightly bound electron, usually from $K$-
or $L$-shell of the atom. Unlike the gradual rate of energy loss of charged particles, the photon's energy transfer to electron happens
abruptly.
The photoelectron
is ejected from the atom with a kinetic energy, $E_{p.e.}$ which is given by $E_{p.e.}=E_\gamma-E_b$,
where $E_\gamma$ is the energy of the incident photon and $E_b$ is the binding energy of the
electron in its original shell. Each photoelectric interaction leaves the atom
in an ionized state with an inner shell vacancy. The de-excitation may take place
by the capture of a free electron from the medium or by a series of internal
electronic rearrangements.
Internal transitions may cause the emission of one or several fluorescent X-rays.
These X-rays can be reabsorbed in close proximity to the original
photoelectric event by interacting with electrons that are bound less tightly.
However, gasses of a counter usually have a relatively low absorption
cross-section for such fluorescent radiation.
This leads to the escape of $K$-fluorescence X-rays from the active volume of the detector.
In this case, the energy deposit is equal to the difference between the incident photon energy and the escaping photon energy.
Background gamma-rays produce Kr $K_\alpha$ X-rays with the energy of 12.6 keV.
The cross section of the photoelectric effect on the $K$-shell is sufficiently large.
This argument is supported by the peak at the energy of $\sim34$ keV, corresponding to the escape peak of 46.5 keV line for krypton
($E_{\gamma46.5}-E_{K_{\alpha \mathrm{Kr}}}=33.9$ keV).
As an example, Fig.\ref{fig:SpcE_81Kr}(b) shows the pulse height spectrum of two-point events with one subpulse being in the energy range
$E=K_\alpha({\rm Kr}) \pm \Delta E$
or $E=K_\alpha({\rm Br}) \pm \Delta E$ for two filling samples of the counter with the krypton enriched in $^{78}$Kr (dark line) and with
depleted krypton containing $^{81}$Kr (shaded region).
We can see a good agreement of background measurements above 20 keV.

The prominent 13.5 keV peak in Fig.\ref{fig:SpcE_81Kr}(b) is associated with the separation of subpulses from the Br $K_{\alpha}$ X-ray and
Auger
electrons from the daughter atom [$K_{\alpha}\otimes e_A$]. The difference between the Kr $K_\alpha$ and Br $K_\alpha$ is $\sim600$ eV.
The relaxation of the $K$-vacancy in the bromine atom in
63.6\%
of cases \cite{Han2007} results in the emission of characteristic X-rays of the following
energies (relative intensities): $K_{\alpha1}=11.924$~keV (100\%);
$K_{\alpha2}=11.878$ keV (52\%);
$K_{\beta1}=13.291$ keV (14\%);
$K_{\beta2}=13.469$ keV (1\%)
$K_{\beta3}=13.284$ keV (7\%)~\cite{2K-12}
and accompanying Auger
electrons of 1.58, 1.62, 0.20, 0.03 and 0.21 keV energies, respectively.
In the case of the X-ray emission, the energy of Auger electrons is enough
to create a cluster of primary ionization.
As a result, a distinguished two-point event is created.
In 38.2\%
of cases, the $K$-vacancy in bromine is filled by the emission of Auger electrons only.
This results in the single-point event.
Single-point events are also created when $K_{\beta1}$ and $K_{\beta2}$ X-rays are emitted.
This happens because Auger electrons then
have the energy that is smaller than the charge-sensitive amplifier's noise. Taking
account of the efficiency of the absorption of the characteristic quanta in
the operating volume ($\varepsilon_{abs}=0.83$), one can estimate
a composition of the total absorption peak of 13.5 keV (TAP): 44.4\%
of single-point events
and 45.0\%
of two-point events
(or 0.894) with respect to the total number of $K$-captures.
As a result of the merge of two sub-impulses into one with probability $P_{m}$,
some of two-point events will be converted to one-point events and,  in fact, there will be detected $(44.4+P_{m}\times45.0)$\%
of single-point and ($(1-P_{m})\times45.0$\%
of two-point events in TAP.
Thus, there will be $(49.7+P_{m}\times50.3)$\%
of single-point and $(1-P_{m})\times50.3$\%
of two-point events in the peak of full absorption, normalized to the area.

Experimental studies of the shake-off process allow us to clarify the efficiency of the event selection.
A background can be minimized by detecting the ejected electrons in coincidence with at least one Br $K$ X-ray.
This X-ray may indicate that the capture event has happened.
Furthermore, decays followed by the double $K$ Auger-electron emission are thus excluded from the measurements, reducing the energy
uncertainty.
Ideally, one might wish to select coincidences with double $K$ X-rays only, which indicates that a completely empty $K$-shell has been filled
by two radiative transitions.
However, the counting rate would be very much reduced by this requirement.
Nevertheless, we can yet specify more precisely the curves of calibration and resolution of energy, after determining the parameters of the
response function of the pointwise charge cluster in the compound event.

\subsubsection{Background processes}

Let us consider background processes that could mimic the
signal from the coincidence $(K_{\alpha}^h \otimes K_{\alpha}^s \otimes e_A)$ in
the experiment and analysis of data.
We will explain below how these contributions can be eliminated from the calculation of
the probability $P_{KK}$ of the double $K$-shell vacancy production per a $K$-electron capture.

\begin{figure*}
\begin{centering}
\includegraphics*[width=2.75in,angle=270.]{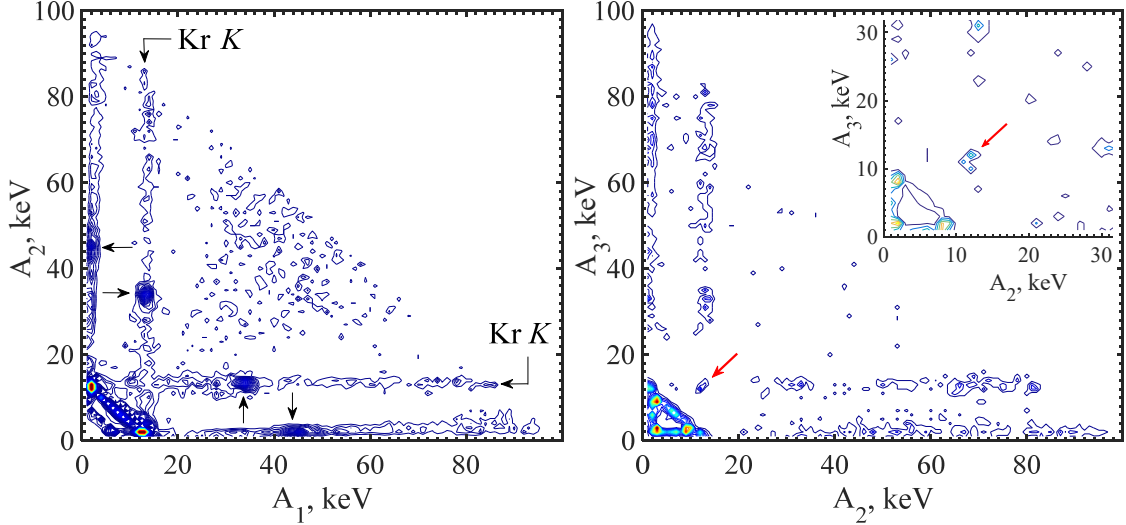}
\end{centering}
\caption{\label{fig:matr2_81Kr}
Two-dimensional amplitude distributions of energy deposits of individual components for
two-point (left panel) and three-point (right panel) events from
the background of the LPC filled with depleted krypton.
In incut of the right panel is demonstrated the same for the enriched  krypton sample.
The arrows indicate to the response from interaction of a 46 keV photons with electrons of krypton atoms.
}
\end{figure*}

The two-dimensional distribution of two-point events
of the total energy of 1-100 keV
with amplitudes (the area under a single gaussian curve in keV) $A1$ and $A2$ of the first (in time)
and the second component of the current pulses is shown in Fig.~\ref{fig:matr2_81Kr} (left).
A flat spectrum of the background, representing a random coincidence of incoherent scattering and bremsstrahlung, was subtracted from the
distributions.
The horizontal and vertical loci of Fig.~\ref{fig:matr2_81Kr} represent the effect of the interaction of background photons with the internal
working matter of the detector and its walls.
The diagonal locus stands for the Compton scattering in
gas and emission of the bremsstrahlung.
One can see that among two-point events there are clearly marked loci in the energy regions of $\sim 1.6$ and $\sim 11.9$  keV, corresponding
to Auger electrons and Br $K_\alpha$ lines.
The corresponding energy spectrum of individual ionization clusters for two-point events with the total $(A1 + A2)$ amplitude up to 18 keV is
shown in Fig.~\ref{fig:spc_Res_Kr81}(a) (black line).
In this figure, we can clearly see the peaks of the Auger electrons and $K_\alpha$ X-rays that appear after the single $K$-capture decay of
$^{81}$Kr (dark line).
The blue dotted curve in the same figure is the energy spectrum of
two-point events in the LPC, filled with the krypton enriched in $^{78}$Kr.
The peak of krypton's characteristic $K$-photons is well visible at the energy of 12.6 keV.

We can select three-point events  with a total energy of 5 to 100 keV.
One of the components there corresponds to the energy release being in the range from 0.6 to 8.5 keV.
For such events, Fig.~\ref{fig:matr2_81Kr} (right), the two-dimensional distribution of the energy release of two (out of three) pointwise
charge clusters with the amplitude $A2$ of one pulse and $A3$ of another one, placed on X- and Y-axis, respectively.
This figure clearly shows loci corresponding to the $K$-shell photoeffect of the krypton atom [$K_\alpha \otimes (e_A+e_{p.e.})$].
In 86.7\%
of cases~\cite{Storm} we dealt with the photoeffect of $K$-electrons.
The fluorescence yield of the $K$-vacancy in krypton is $\omega_{K}=0.660$, and Auger electrons are emitted in 34\%
of cases~\cite{2K-12}. This means that
57.2\%
of photoelectric absorptions can create two clusters of primary ionization.
Another part of photoabsorption (42.8\%)
will generate a single cluster of an ionization by Auger and photoelectrons (also from $L-$, $M-$, $N$-shells).
Loci around $\sim 1$ and $\sim 8$ keV are clearly visible in Fig.~\ref{fig:matr2_81Kr} (right panel).
These loci are formed by  coincident responses from the krypton Auger electrons
and the copper $K$-fluorescence excited by the Kr $K$ X-rays on the
copper wall of the detector.
These events can be considered as a closely spaced Gaussian centroid that is typical for near-wall events.
The parameters of such events depend on the time that it takes for the primary
charge cluster to drift towards  the anode.
As it drifts, the charge cluster spreads out into a cloud due to the electron diffusion.

As noted above, a double $K$-vacancy in the daughter atom can appear after a single $K$-capture
in the $^{81}$Kr, which is present in the depleted sample.
This vacancy can be created by a sudden change in the nuclear charge $Z\rightarrow(Z-1)$.
As a result, another $K$-electron can be excited to a higher state in the SU process or it can be completely removed from the electron shell in
the SO
process~\cite{Freedman_74}.
The SU process contribution is relatively
small. The energy balance of the SO process is as following:
\begin{equation}
Q_{0}=[T_{e}+2(B_{K}^{'}-X_{K}^{'})]+T_{\nu}+2X_{K}^{'},
\end{equation}
where $Q_{0}$ is the total transition energy (atomic mass difference);
$T_{e}$ is the kinetic energy of the released electron; $B_{K}^{'}$
is the binding energy of the $K$-electron in the daughter atom; $X_{K}^{'}$
is the energy of X-rays of the daughter atom; $T_{\nu}$ is the kinetic
energy of the neutrino~\cite{rare279}. In this equation, the shift
of $B_{K}^{'}$ value in the atom with the two $K$-vacancies with respect to the value for the atom with the one $K$-vacancy is not taken into
account.
The maximum kinetic energy of the electron is $T_{e}^{max}=Q_{0}-2B_{K}^{'}$.

The probability creation of  a double $K$-vacancy as a result of a single $K$-capture was measured for fifteen isotopes~\cite{rare279,rare280,rare281,rare282,rare283}.
Table IV in the publication of Hindi
and Kozub \cite{PhysRevC.45.1070} provides us with an overview of the measured values for
$P_{KK}$ in $EC$ decays.
The dependence of the probability of creation of the double $K$-vacancy
per one $K$-electron capture can be described~\cite{rare280} with a
moderate accuracy as:
\begin{equation}
\label{P_KK_apr}
P_{KK}\approx P_{SO}\approx0.08\cdot Z^{-2}.
\end{equation}

The double-vacancy production that follows a $K$-shell electron capture in $^{55}$Fe is the most successful and thoroughly studied process.
The total yield of the $KK$-ionization reached  $(1.53 \pm 0.08) \times 10^{-4}$  for a single electron-capture of $^{55}$Fe in work
\cite{PhysRevC.89.014609}.
The theoretical description of electron ejection during $EC$ decays was developed in Refs.~\cite{Primakoff,Pengra63,Chon94}.
The theoretical probability for the electron ejection is small and falls off steeply with the energy increase.
The energy spectrum of $K$ electrons ejected from $^{55}$Fe during the $K$-capture decay has been measured with a good accuracy in
Ref.~\cite{rare279}.
The predicted values appear to be in 5-180 keV range.
General characteristics of the spectrum of SO electrons that were ejected
from the bromine $K$-shell are similar to those measured for $^{55}$Fe
but with a correction to the transition energy: $Q_{0}=232$ keV for $^{55}$Fe,
$Q_{0}=281$ keV for $^{81}$Kr~\cite{2K-25-1}.
About 80\%
of ejected $K$-electrons have energy in the range of 0-10~keV, including 28\%
- in the 0-2 keV interval.

In our analysis of the accumulated data, the double $K$-shell ionization of
the $^{81}$Br daughter was studied by triple coincidence measurements between
two practically simultaneous radiations resulting from the filling of the two holes in the $K$ shell:
hypersatellite- and satellite-line photons and bromine emitted electrons
(sum Auger electrons and $K$ electron ejected) -  $[K^h \otimes K^s \otimes (e_{A}+e^K_{\rm{SO}})]$.

Energy and other parameters of the hypersatellite and the satellite X-ray, emitted during the filling of either of two vacancies of the
daughter bromine atom, were estimated in the framework of the Dirac-Fock method, using the RAINE package.
The energy shift of Br $K_\alpha^h$ hypersatellite with respect to the diagram lines has approximately accounted for 370 eV.
The energy shift of Br $K_\alpha^s$ satellite with respect to diagram lines have amounted to several tens of eV.
The results of our calculations are presented in Table~\ref{tab:Calculated-and-reference}.
The energy shift of these $K_{\alpha}^h$ hypersatellites with respect to the  diagram lines has been calculated by several groups
\cite{Desclaux, Beatham, MCDF}, including the splitting of the final $KL$-double-hole state due to the Breit interaction.
The average Breit energy of the configuration was calculated in the long-wavelength approximation.
Results of our calculations are in good agreement with theoretical predictions of Ref.~\cite{PhysRevA.25.391}.
Then the total energy is
\[E=2B_{K}^{'}+E_{\Delta}+T_{e}=27470~\texttt{eV}+T_e.\]

As far as we know, no measurements exist for the fluorescence
yields of the $K$-shell vacancy filling in the presence of an additional vacancy.
In all previous measurements of $P_{KK}$, it was assumed that the additional vacancy does not affect the fluorescence yield of the $K$-vacancy
filling.
In work~\cite{2K-25}, M.H.~Chen made the calculation of
Auger rates, radiative transition rate, and fluorescence yields of
the double-$K$-hole state using the MCDF method.
According to this work, the fluorescence yield $\omega_{2K}$ from the $2K$-vacancy for elements
with $Z=36$ is, at most, 2.6\%
larger than the yield from the single $K$-vacancy.

From what is mentioned above, we can assume the following variant of radiation composition
(in addition to $e_{SO}^{K}$):
$a$) $e_{A}^{K} \otimes e_{A}^{K} \otimes e_{A}^{L}$; $b$) $X \otimes e_{A}^{K} \otimes e_{A}^{L}$;
$c$) $X \otimes X \otimes e_{A}^{L}$ with the probabilities of being 0.124, 0.464, 0.412, respectively.

This composition  will contribute to the TAP at the level of 100\%
in $(a)$, 83\%
in $(b)$ and 68.9\%
in $(c)$ because of the escape of the X-ray from the detector.
Because of the overlap of subpulses, $\sim$0.255 events of type $(b)$ will become the single-point ones, $\sim$0.065 and $\sim$0.38 events of type $(c)$ will
become single- and two-point ones, respectively.
As a result, the TAP will contain
0.26 of the single-point, 0.392 of the two-point, 0.148 of the three-point
events in the initial number of $2K$-vacancies.

Groups of subpulses of the hypersatellite $K_{\alpha(\beta)}^{h}$(Br) [subpulse $A_x$], satellite $K_{\alpha(\beta)}^{s}$(Br) [ subpulse
$A_y$],
and of the bromine emitted electrons $(e_{A}^{L}\oplus e^K_{SO})$ [subpulse $A_z$] will be among the required ``useful'' three-point
events.
We will assume
that the filling of outer ($L, M, N$)-shell vacancies will produce Auger
electrons with the total energy of $(B_{2K}-K_{\alpha(\beta)}^{h}-K_{\alpha(\beta)}^{s})\sim3.3$ keV.
The spectrum of $(e_{A}^{L}\oplus e_{SO})$ is limited at low energies due to the finite energy of Auger electrons.
The background for our search of useful events will be greatly reduced if the magnitude of the subpulse $A_z$ is set to be in the range
3.3-10.3 keV.

\subsubsection{Experimental results}

The total number of $^{81}$Kr $K$-captures can be estimated from the area under the 13.5 keV TAP curve for all types of events in
Fig.~\ref{fig:SpcE_81Kr}(a) as
\begin{equation}
\label{eq_N_K}
  N_{K}= \sum\limits_i {\frac{N^{exp}_{K}(i)} {\epsilon _{d}(i) }}=7.9\times 10^6,
\end{equation}
where
$N^{exp}_{K}(i)$-the number of events that belong to the energy range 10.5-16.5 keV for a single stage,
$\epsilon_{d}(i)$ is the absolute efficiency to detect the respective radiation for  $i$=[\emph{I}, \emph{II}, \emph{III}]-stage number.
Such statistics is enough to detect satellite- and hypersatellite-line photons generated during the shake-off process in the LPC.
With this in mind, we analyzed pointwise charge clusters of the three-point events data.

Figure \ref{fig:matr2_81Kr} (right) displays a clearly visible locus in the ($A_2\sim 12$ and $A_3 \sim 12$ keV) region where the LPC is
filled
with depleted and enriched (in the inserted panel) krypton gas samples.
The events at that locus of depleted krypton (containing $^{81}$Kr) can be generated by the coincidence of $K^h$ and $K^s$ X-rays.
These rays are formed in the process of filling of the double $K$-shell vacancy.
They are formed in both a single $K$-electron capture and when a single photon is absorbed with releasing both $K$ electrons and creates a
``hollow atom''.

In the case of the enriched sample, the locus may include the events of the coincidence of
$K^h$  and $K^s$ X-rays,
produced by  $2\nu2K$ decay of $^{78}$Kr.
The value of the subpulse $A_z$ will be 1-4 keV
for the enriched sample and 1-10.3 keV for the depleted sample because the $K$-shell shake-off electron,
which follows the $K$-capture, has a broader energy distribution.

Summing the number of triple $[K^h \otimes K^s \otimes (e_A+e^K_{SO})]$ coincidences  for each  stage of the data set,  we can
obtain a
statistically significant value for the double $K$-shell ionization during the $EC$ decay of $^{81}$Kr.
Similarly, summations were carried out for the spectra of $K_\alpha$ and $K_\beta$ normal diagram lines in bromine
to obtain the integrated intensity $N_{K}$ for the predominant single $K$-vacancy transitions.

\begin{figure}
\begin{centering}
\includegraphics*[width=2.75in,angle=0.]{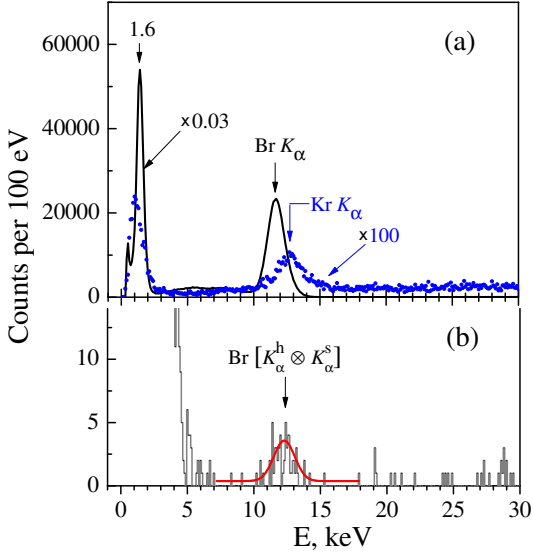}
\par\end{centering}
\caption{\label{fig:spc_Res_Kr81}
(a) the spectrum Auger electrons (peak 1.6 keV) and $K_\alpha$ after the single $K$-capture decay of $^{81}$Kr and (b) a fragment of the
coincidence spectrum with [$K^h \otimes K^s$].}
\end{figure}

Figure \ref{fig:spc_Res_Kr81}(b) shows the magnitude of the coincidence in the $K^h$-$K^s$ region of the two pointwise charges clusters of
three-point events under the condition that the third is in the
range from 0.6 to 8.5 keV.
As it was expected, the peak in the region of $\sim 12$ keV is produced by the simultaneous detection of the Br $K^h$ hypersatellite and $K^s$
satellite X-rays.

The total number of triple coincidence events $[K\otimes K\otimes e_{SO}]$ in the operating volume of the LPC, which is expected to be recorded
over the full time of data collection, can be defined as
\begin{equation}
\label{eq=Nkk}
N_{KK} = N_K P_{KK} \omega^h  \omega^s \delta _e \eta = 56\pm8,
\end{equation}
where $N_K$ is the number of $K$-capture during $^{81}$Kr decays,
the probability of the double $K$-shell vacancy production per $K$-electron capture in $^{81}$Kr is $P_{KK}=6.5\times10^{-5}$.
It is calculated according to Eq.~(\ref{P_KK_apr}).
In addition, $\delta _e=0.7$ is the fraction of all electrons, that is shaken into the continuum,
registered in the coincidence according to the selection criteria and
$\eta$ is the complete coefficient of selection for the $K^h$ and $K^s$ X-rays.
This coefficient is calculated as $\eta = \varepsilon_p  \varepsilon_3  \alpha_k $,
with $\varepsilon_p =0.81 \pm 0.01$ being the probability of two $K$ photons to be absorbed in the operating volume;
$\epsilon_{3}=0.54 \pm 0.06$ being the efficiency to select three-point events;
$\alpha_{k}=0.985 \pm 0.005$ being the fraction of events with two $K$ photons that could be registered as distinct three-point events.

At the same time, the total number of events that have formed the locus ($A_2\sim$ $A_3$~$ \sim 12$ keV) amounted to $N_{coinc}^{dipl}=42 \pm
7$.
Events in the locus can include signals from both, $K$ shake-off process and $K$-shell photoionization of an atom by a background photon.
Although we did not distinguish between these processes in our experiment, the contribution from the double $K$-shell photoionization of the
atom is negligible.
It is because the cross section of the photoionization of the $K$-shell tends to be zero near the threshold.

The obtained value $N_{coinc}^{dipl}$ is a satisfactory agreement with the estimate~(\ref{eq=Nkk}).
We assumed that the number of expected  background events in the locus was no more than one event.
According to the Feldman-Cousin procedure~\cite{Feldman98}, the maximum value of the background in the energy region of our interest is 1.6
events with a confidence level of 90\%
for the full measurement time.

In order to obtain the probability of double $K$-shell vacancy production per $K$-electron capture for the shake-off process we rewrite formula (\ref{eq=Nkk}) and find
\begin{eqnarray*}
  P_{KK}^{SO} &=& \frac{1} {N_{K}} \times  \left[\frac{N_{coinc}^{dipl}} {\omega^h \omega^s \delta _e \eta}\right]
   = \nonumber       \\
    &=& [5.4 \pm 0.8 (stat) \pm 0.4 (syst)] \times 10^{-5}.  \nonumber
\end{eqnarray*}
The systematic error includes the uncertainties due to the data acquisition and handling.
This value is in a good agreement with the theoretical predictions of Suzuki and Law \cite{Suzuki1982}.
Based on the theoretical calculations of Ref.~\cite{Suzuki1982}, the probability for the shake-up is much smaller than the probability for
shake-off, giving $P_{KK}^{exp} \cong P_{KK}^{SO}$.

\subsection{Double $K$-capture of $^{78}$Kr}

Based on the above-mentioned assumptions, we can postulate that the shake-off process is responsible for all events in the locus
($A_2\sim$~$A_3$~$ \sim 12$ keV) of the depleted krypton sample.
In addition, the contribution of random coincident background events is less than the statistical error.
At the same time, the total number of events in the same locus was estimated to be
$N_{coin}^{enr}=16 \pm 4$ after reanalyse 782 live days of data collected with the enriched sample.
In contrast to the previous analysis \cite{Gavr2013PRC,PAN2013}, the time dependencies of the spectrometric properties
of the LPC were taken into account here, as in the case of a depleted sample.
The magnitude of the background under the locus we evaluate as close to one event.
This magnitude was estimated using the surrounding neighborhood of the locus of our interest and taking into account the measurements of
background with the sample of depleted krypton.
Assuming this effect to be attributed to the $2\nu2K$ capture in $^{78}$Kr, one can estimate the half-life of this process at 90\%
C.L.:
\begin{eqnarray*}
  T_{1/2}^{2\nu 2K}  &=& ln2 N_A \times \frac{\omega_{2K}\cdot  \epsilon_{f} \cdot t } {N_{coin}^{enr}} =\\
   &=& [1.9^{+1.3}_{-0.7}(stat) \pm 0.3 (syst)] \times 10^{22} \texttt{yr},
\end{eqnarray*}
where $N_A=1.08\times10^{24}$ is the number of $^{78}$Kr atoms in
the fiducial volume of the counter,
$\omega_{2K} = \omega^h \times \omega^s=0.47$ is the fraction of $2K$-captures accompanied by the emission of
two $K$ photons, $\epsilon_{f}$
is the detection efficiency
and $t$ is the total live time of the experiment.
The efficiency $\epsilon_{f}$ was calculated based on the following expression
$\epsilon_{f} = \varepsilon_p  \varepsilon_3  \alpha_k  k_\lambda$,
where $\varepsilon_p$,  $\varepsilon_3$ and $\alpha_k$
as in the Eq.~(\ref{eq=Nkk}).
$k_\lambda$ is a useful event selection coefficient that can defined for a given threshold value of $\lambda$.
This coefficient takes into account edge effects [see Fig.~\ref{spcB_Krn_II}].
The systematic error includes uncertainties of spectroscopy subpulses vs time of measurement.
The largest source of uncertainty is associated with the absolute normalization of $\varepsilon_3$.
Statistical errors are defined according to the recommendations of Ref.~\cite{Feldman98}.

\section{\label{sec4}CONCLUSION}

In this paper, we carried out the comparative study of the signal from the decay of double $K$-shell vacancy production that follows after
single $K$-shell electron capture of $^{81}$Kr and double $K$-shell electron capture of $^{78}$Kr.

The radiative decay of a the double $1s$ vacancy state was identified by detecting the triple coincidence of two $K$  X-rays and several Auger
electrons in the $ECEC$-decay, or by detecting two $K$ X-rays and (Auger electrons + ejected $K$-shell electron) in the $EC$ decay.
The double $K$-vacancy creation probability in the electron capture decay of $^{81}$Kr was measured with a reasonable statistical
and systematic precision.
We have determined $P_{KK}^{exp}$ for the $EC$ decay of $^{81}$K, which are in satisfactory agreement with the value of $P_{KK}$  calculated
expected with  $Z^{-2}$ extrapolation (Eq.~\ref{P_KK_apr}), predicted by Primakoff and Porter \cite{rare280}.
This allowed us to more accurately determine the background in the energy region where the search is carried out $2K$-capture in $^{78}$Kr
using of LPC.
For krypton samples of different enrichment, the general procedure of analysis of LPC data has allowed us to determine the half-live of
$^{78}$K with respect to the $2\nu2K$-to ground state-capture. This determination is more precise than the one of the work \cite{PAN2013}.

\begin{center}
    {\textbf{ACKNOWLEDGMENTS}}
\end{center}

We would like to thank Oleksandr Koshchii (The George Washington University) for
productive discussions and careful reading of the manuscript and useful suggestions, respectively. The work was made in accordance with INR
RAS
and V.N.~Karazin KhNU plans of the Research and Developments.


\end{document}